# Research on Creative Thinking Mode Based on Category Theory

Tong Wang

(Guangzhou College of Technology and Business)

## Abstract

    The research on the brain mechanism of creativity mainly has two aspects, one is the creative thinking process, and the other is the brain structure and functional connection characteristics of highly creative people. The billions of nerve cells in the brain connect and interact with each other. The hundreds of millions of nerve cells in the brain connect and interact with each other. The human brain has a high degree of complexity at the biological level, especially the rational thinking ability of the human brain. Starting from the connection of molecules, cells, neural networks and the neural function structure of the brain, it may be fundamentally impossible to study the rational thinking mode of human beings. Human's rational thinking mode has a high degree of freedom and transcendence, and such problems cannot be expected to be studied by elaborating the realization of the nervous system. The rational thinking of the brain is mainly based on the structured thinking mode, and the structured thinking mode shows the great scientific power. This paper studies the theoretical model of innovative thinking based on of category theory, and analyzes the creation process of two scientific theories which are landmarks in the history of science, and provides an intuitive, clear interpretation model and rigorous mathematical argument for the creative thinking. The structured thinking way have great revelation and help to create new scientific theories.

## I. Introduction

cogito, ergo sum (I think, therefore I am)

    ---René Descartes

    From ancient times to the present, the discussion about human rational thinking has always been one of the philosophy topics. Aristotle put forward formal logic in "On Instruments" and studied the syllogistic rules of propositional reasoning, which is the generalization and summary of human rational thinking by ancient philosophers. Similarly, the discussion of human rational thinking is also at the core of modern philosophy. Philosophers have established various theories of mind in the name of "anthropology" or "human knowledge research", forming two camps of "rationalism" and "empiricism". In "Critique of Pure Reason", Kant proposed the cognitive model of rationalism and empiricism---transcendental idealism and rational intuition. Russell was extremely critical of Kant's philosophy, believing that human rational thinking lies in free creation and does not recognize the presupposition of transcendental idealism. Scientists in modern philosophy and cognitive research believe or assume that human beings have a certain innate cognitive structure of the brain. By studying and understanding this cognitive structure, we can solve the problem that how rational human cognition is possible.

    The development of modern technical means has made great progress in the study of the brain's thinking mode. For example, functional magnetic resonance imaging technology provides technical support for revealing the creative brain structure and functional network connections [1], providing people to study the brain structure and connection with function. Brain imaging technology is used to study the unique phenomena such as epiphany and the internal connection of the brain [2]. Based

Tong Wang: wangtong_phd@163.com

on electroencephalogram (EEG) and brain functional imaging research technology, Psychologists directly observe the activity of the brain when processing complex information, so as to explore the brain mechanism of creative thinking (Bowden, Jung-Beeman, & Zf, 2007; Fink et al., 2007; Luo & Knoblich, 2007; Srinivasan, 2007). So far, there are two main aspects of brain mechanism research on creativity, one is the creative thinking process, and the other is the brain structure and functional connection characteristics in highly creative people [3] [4] [5].

Creativity is the advanced manifestation of human rational thinking, the glory of human mind, and the source of the progress of human civilization. However, creative thinking itself is complex. As is seen in above literature, with different experimental conditions and technical means, the research results of the brain mechanism of creative thinking are very different. It is therefore difficult to have a clear and deterministic conclusion about the brain mechanisms of creative thinking (one-to-one explanatory model). No studies have shown that creative brain functional connectivity can be altered by training and then have practical meaning.

Hundreds of millions of nerve cells in the brain are interconnected and interact with each other, producing people's superb motor movements, and also emerging people's complex advanced cognitive activities, such as perception, language, thinking, etc. The human brain has a high level of biological complexity, especially the rational thinking ability of the human brain. Research in brain science involves neuroanatomy, neurophysiology, neural networks, neurolinguistics, and more. Starting from the connection of molecules, cells, neural networks and the neural function structure of the brain, it may be fundamentally infeasible to study the rational thinking mode of human beings. Human's rational thinking mode has a high degree of freedom and transcendence, and such problems cannot be expected to be studied by elaborating the realization of the nervous system.

The rational thinking of the brain is mainly based on the structured thinking mode. The structured thinking mode of the brain shows the great scientific power, and this thinking mode is the main mode which had the major scientific discoveries in the whole mankind history. This paper studies the structural characteristics of the brain's thinking mode based on the category theory, and the brain's structural thinking mode is clearly explained and understood. Understanding and mastering the structured way of thinking has great enlightenment and help to create new scientific theories.

## 2. Category Theory

2.1. History and Definition of Category Theory

As early as in ancient Greece, Aristotle wrote his famous "Categories", which explored the classification of objects that can be recognized by humans. In the context of modern mathematics, categories have other meanings and precise mathematical definitions. But, the basic intuition remains unchanged: a category is formed by grouping together a class of similar objects. Category theory was founded in MacLane and Eilenberg's 1945 paper "General theory of natural equivalences" [6]. Category theory has expanded into most fields of modern mathematics in a very short period of time, and category theory is the language and thinking way of describing abstract mathematical structures.

Definition 1: A category C includes:

(1) A collection of a set of objects O; as a mathematical object, it has internal operation rules;

(2) A collection of morphisms, in which morphisms (or arrows, functors f) f: $X \to Y$; X is the domain of f, and Y is the codomain, denoted as $\text{dom}(f) = X, \text{cod}(f) = Y$. $\text{Hom}(X,Y)$ to denote the entirety of arrows from object X to Y. This is a collection, as shown in Figure 1(a) below.

For any object X, there is an arrow, as shown in Figure 1(b) below, $1X: X \rightarrow X$, 1X is named the identity of X, which is the unit law. For any f, $f \circ 1X = f = 1Y \circ f$. There are two arrows f and g, such that $cod(f) = dom(g)$, as shown in Figure 1(c) below. $g \circ f$, denote the composition of f and g. A graph is said to be "commutative" when any arrow composite has the same range and equal domains. As shown in Figure 1(d) below.

$$X \xrightarrow{f} Y \quad X \xrightarrow{1X} X \quad Z \xleftarrow{g} Y \xleftarrow{f} X \quad Z \xrightarrow{g \circ f} X$$

(a)      (b)      (c)      (d)

Figure 1   Definition of category

With functors as morphisms, all categories form a larger Cat. So naturally, we have the notion of isomorphism between categories: as isomorphisms within other categories, two categories X, Y are isomorphic if and only if there are two functors F: C→D, G:D→C, so that both sides are identity functors after they are combined, as shown in Figure 2 below:

(3) $G \circ F = idC, F \circ G = idD$

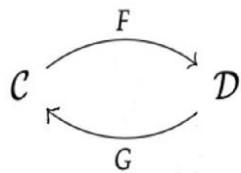

Figure2 Category isomorphism

The concept of "isomorphism" is defined by a compositional functor, a thinking way that is consistent with the philosophy of category theory, which defines mathematical structures in terms of morphisms. In other words, the basic ideas of category theory reflect the way we organize the structure of information. The idea that morphism represent category has been widely used in the foundations of mathematics. For example, a subgroup can be equivalently regarded as a single group homomorphism, and the quotient group (or normal subgroup) of a group can be equivalently regarded as a full group homomorphism; thus, a subset can be equivalently defined in terms of injective, a quotient can be equivalently expressed in terms of surjective. Developing this idea, the concept of homology which measuring the difference between the exact sequence and the surjective of the category can be defined.

From the above definition, the basic idea of category theory is to study the properties of an object through the relationship between objects. This relationship is usually defined by morphism, so that morphism reflects the structure of mathematical objects. "Structure" refers to a class of mathematical objects described by axioms, and mathematics is concerned with "structure" rather than a specific "object". Bourbaki believe that the structure in mathematics is divided into three categories: algebraic structure (group, ring, field), order structure (partial order, total order) and topological structure (limit, continuity, connectivity, neighborhood). With the improvement of the abstraction degree and the expansion of the field of mathematics, the structure itself is the "category theory" as a study object.

Category theory is the study of mathematical structures in a concise, general and abstract way. For example, in topology, a doughnut is the same as a coffee cup. According to category theory, they are isomorphic in a category Top, the object is a topological space, and the arrow is a continuous map (i.e., continuous transformation), so a doughnut and a coffee cup are homeomorphic.

The advantages of category theory are as follows:

A very obvious trend in the development of modern science is the division between disciplines; knowledge in different fields seems to be divided more and more finely, and it is more and more difficult for us to have an overall understanding of it. At the conceptual level, category theory unifies definitions and concepts from different branches of mathematics. Category theory has partially unified the division of mathematics at the conceptual level, and it has found the same conceptual basis among different branches of mathematics.

It's promising to study difficult interdisciplinary scientific problems applying the structural ideas of category theory. For example, physicists apply category theory to solve the problem of topological condensed matter. A good study of problems in mathematical logic. Applying the structured thinking mode of category theory, the structured thinking mode of the brain is researched, and through the Bi-interpretation perspective of the category theory, the structural characteristics of the brain's thinking is interpreted.

## 3. Category Theory Model of Structured Thinking

3.1 Transcendence of creative thinking

As early as in ancient Greece, the study of human rational thinking has already begun and established. Plato vividly expresses his philosophical "idea" view through the cave metaphor, in which the "idea" is seen as a being that illuminates everything. And to reach knowledge of this idea, mind must go through a series of turns, from shadows to things, to firelight, to things outside the cave, and finally to the sun. The path of individual liberation from opinions and phenomena is the same path [7].

Parmenides was the first to put forward the schema of "thought" (rational thinking) which is opposite to feeling. This is the first real thinking model in the study of human reason. Later philosophers including Kant modified and developed this model.

As shown in Figure 3, Parmenides classified objective matter and subjective spirit, and proposed two philosophical categories of "existence" and "non-existence", "feeling" and "thought". In it, what can be "thought is the same as that which can be", and feeling and non-being are the same. Parmenides gave a negative answer to the question of whether there is a connection between existence and non-existence, feeling and thought. Rational cognition leads to the "path of truth", while perceptual cognition leads to It is the "opinion road", the two roads are diametrically opposed [8].

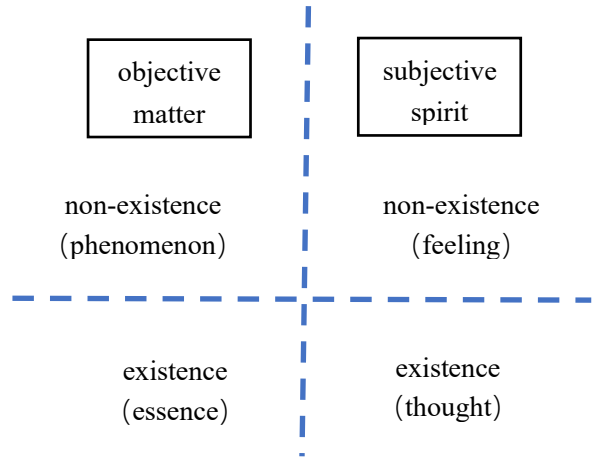

Figure3  Parmenides' model of rational thinking

After the Renaissance, modern philosophy, physics, mathematics, and neuroscience have developed by leaps. The cognition of rational thinking has undergone a huge change, and Descartes put forward the famous slogan "cogito, ergo sum (I think, therefore I am)", which is logically impossible to refute, that is, all the world of experience, all senses and all idea are some kinds of illusion. But the existence of this activity itself indicates a being, which is "I", and rational cognition of the world can only proceed from doubt. From the perspective of human philosophy, Cassirer, modern philosopher and scientist, regards man as a symbolic animal, and only in man, the question of possibility and reality can be realized. Only in man, mythical, religious, linguistic, artistic, historical and scientific symbols can be created and applied. Man use symbols to create culture, and the creation and using of symbols is the difference between humans and animals [9].

In the 20th century, with the establishment of the two major scientific theories of relativity and quantum mechanics, people have a deeper understanding of the characteristics and structure of scientific theories. Now generally, it is believed that the structure of a scientific theory consists of three elements: concepts; basic principles, linking to these concepts; the logical conclusions, i.e., various specific special laws and predictions derived from these concepts and principles. Among them, basic concepts, basic principles and propositions constitute the core elements of theory.

Einstein said, "in the structure of my thinking, written or spoken words do not seem to have any role... The various concepts that appear in our thinking and our verbal expressions are logically free creations of the mind, they cannot be derived inductively from sensory experience. The reason this is not so easy to notice is only because we are so accustomed to associating certain concepts and their relations (propositions) with certain sensory experiences in such certainty that we are unaware that such a logical an insurmountable chasm that separates the world of sensory experience from the world of concepts and propositions" [10].

According to Einstein, "There is nothing that can be said a priori about the formation of concepts and how they are related to one another, and how we oppose these concepts to sensory experience. Comparing these with the rules of the game, in the game, the rules themselves are arbitrary, but the game is possible only if they are strictly followed. But such rules are never final, and they will only be effective if they are obeyed in a specific field"[11].

Reviewing the historical development of rational thinking, it's obviously that the transcendence of brain thinking, beyond the world of experience, and even beyond human intuition, which guided scientists to establish various scientific theories.

3.2 Category Theory and Natural Transformations

In order to study the structural characteristics of brain thinking, it is necessary to further study the structural characteristics of scientific theories, that is, to categorize and axiomize scientific theories. In this way, from the perspective of categorization, the isomorphic characteristics of theoretical models of different disciplines and the categorization characteristics of brain thinking can be shown.

Definition 2:

A scientific theory category C includes:

(1) A collection T of a set of objects; the set contains an operator O and its operation rules;

(2) A collection of morphisms, in which morphisms (or arrows, functors f) f: T → T ;  f is said to be an automorphic functor induced by object collection. Let $\text{Hom}(T, T)$ denote the entirety of arrows from object T to T.

Definition 3:

A scientific theory is a $\text{Hom}(T, T)$.

The conception of scientific theory is defined through the predicate "is a $\text{Hom}(T, T)$". As far as scientific theory is concerned, the predicate $\text{Hom}(T, T)$ represents a mathematical structure, which is the basic structure of scientific theory. If only the mathematical structure of this theory is considered, this basic structure is sufficient to explain the characteristics of the theory itself.

3.3 The Category Theory Model of Structured Thinking

Definition 4:

F, G: A → B is a functor, a natural transformation, or a morphism t: F → G, is collection of $t_x: x \in \text{Ob}(A)$ such that:

(1) $t_x: F(x) \to G(x)$ is a morphism in category B, and

(2) For each morphism in A (internal operation in A) ϕ: x → y , follow the commutative diagram is shown in Figure 4:

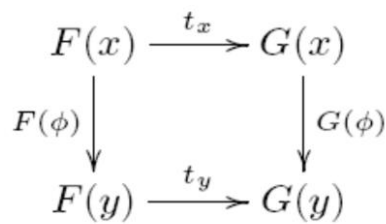 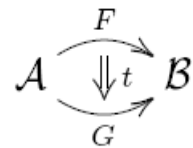

Figure 4  natural transformation　　Figure 5 simplified commutative diagram

Sometimes, commutative diagram is simplified as shown in Figure 5, indicating that t: F → G morphism.

The morphism between functors is named the natural transformation of the functor. There is a mathematical structure of natural transformations between functors, that is, a set of objects, morphisms, and morphisms between morphisms, which is called a 2-category. A general category is called a first-order category. For any category K of order 2, there is also a natural equivalent notion that is weaker than isomorphism. For two objects X, Y in any K, the two objects are equivalent in K, if the composite map of two morphisms f: X → Y, g: Y → X  and the identical map meet the demand of definition of higher-order isomorphism.

3.3 The Category Theory Model of Structured Thinking

By investigating the process of scientific theory discovery, the category theory model of structured thinking of scientific theory is to be established. The father of modern science, Galileo, faced a lot of challenging problems when he created new physics, such as the following:

1. How to find simple laws from the complex appearances, and what principles and tools can be based on? 2. Compared with the old theories, how can the proposed laws make breakthroughs in the interpretation of experience? 3. What kind of mathematical properties (structural) should the found laws have?

The first question, Galileo, combined with his own study of motion, believed that the natural world itself is simple and harmonious. The language of mathematics is simple and clear, and scientific research should follow this language to explore the secrets of nature. This grand book (universe) is written in the mathematical language, and its characters are triangles, circles or other geometric figures, without which it is impossible to humanly understand a word; without these one is wandering in a dark labyrinth [7].

The second question, Aristotle's theory is closely integrated with people's daily experience. Compared with Galileo's new theory, we even can say that Aristotle's theory has more support from empirical evidence of various motions on the earth's surface. Galileo's argument revealed that existing experience must be reinterpreted critically, thus allowing his theory to be reborn from the new interpretation. Galileo's interpretation also takes man out of the Aristotle's cave of sensory experience.

The third question is very crucial, even determining whether the new theory can become a scientific theory. To answer this question, Galileo proposed the principle of inertia (relativity) principle and Galileo transformation. The principle of inertia became the foundation of physics, and this principle evolved into the basic idea of modern physics: Symmetry determines the equations of physics.

The $Ox'y'z'$ coordinate system moves at a constant speed u relative to the $Oxyz$ system. When $t=t'=0$, $O$ coincides with $O'$. The space-time correspondence between the two systems is

$$\begin{cases} t = t' \\ r = r' + ut' \\ v = v' + u \end{cases} \quad (1)$$

This is the Galileo transformation. The transformation is written in matrix T, T connects the two systems and keeps the corresponding physical phenomena consistent in the two systems, which is the isomorphism of the two systems, see Definition 4 and Figure 4.

Similarly, if we propose a different transformation relation (non-isomorphism functors), then a different scientific theory is proposed. For example, the transformation is Lorentz transformation, then the special relativity theory is proposed. These theories are isomorphic (physically equivalent theories), if the proposed functors have natural transformation which have properties of composite mappings and identity mappings. In Fig. 5, the A coordinate system and the B coordinate system are isomorphic if the transformation is Lorentz transformation, and they are physically equivalent special relativity.

Another example of the pinnacle of human creative thinking is the introduction of general relativity. According to the special theory of relativity, all inertial reference frames that move at a constant speed relative to each other are of equal weight, and the laws of physics have the same form in any inertial reference frame. Should the laws of physics in an accelerating frame of reference also be in the same form as those in an inertial frame of reference? Einstein also often thought about the problem of "gravity ": how to include gravitation in the framework of relativity? [12]

Einstein devised a thought experiment in which a person in an elevator could not tell whether he was ascending at a uniform acceleration or in the gravitational field of the ground. The reason for this phenomenon is that we cannot distinguish gravitational fields from inertial fields by any physical experiment. Gravity is very different from other forces, such as electricity. Because we can't cancel out electricity with something like acceleration! But, why can gravity be eliminated? Perhaps gravity should not be regarded as a force at all, but rather as a property of curved space-time. From the idea that gravity is equivalent to acceleration, another surprising conclusion can be deduced: gravity can be eliminated by choosing an appropriate acceleration reference frame. The idea of gravity as a property of space-time led Einstein to the general theory of relativity [11]. With the equivalence principle, any non-inertial frame can be converted into an inertial frame, as long as the influence of a gravitational field is properly processed.

This inference is crucial in the development of general relativity, and its significance is equivalent to the Lorentz transformation in special relativity. From the viewpoint of equivalence of the second-order category, if the transformation F is decomposed and expressed as $F = L + A$, where L is the Lorentz transformation and A is the additional acceleration vector (or gravitational field acceleration), then $F = L + A$ still holds in Figure 5. The corresponding physical phenomena in the two systems A and B are consistent, that is to say, the two systems are isomorphisms of order 2. in Definition 4. Physically, it means that the laws of physics in the local inertial frame can be extended to the laws of physics in any reference frame through generalized space-time coordinate transformation. Einstein took the equivalence principle as a starting point, and finally established the general relativity following the Poisson equation of the gravitational field and Riemann geometry.

Functors and their natural transformations fully demonstrate that the brain has the transcendence of creative thinking, and such transcendence is manifested everywhere in the process of major scientific discoveries. A scientific theory has its specific mathematical structure, and this mathematical structure originates from the free creation of scientists. In the process of free creation, scientists consciously or unconsciously apply the ideas of category theory. The structured view of category theory simplifies the complex and tentative thinking process, making the thinking process clear and structured. Therefore, category theory is very important in knowledge expression and reasoning. The thinking way based on category theory has gradually become a typical paradigm in the innovative thinking.

## 4. Conclusion

Most of the research on innovative thinking is based on neuroscience. This paper analyzes the limitations of existing research methods from the perspective of category theory, and proposes research model on innovative thinking based on category theory. This paper applies the thinking method of categorical equivalence to analyze the creation process of two landmark scientific theories in the history of science, the case is real, and a clear mathematical argument is provided.

Category theory studies structure which is formed by the relation between objects, so the focus of research lies in the relation between objects. Concepts in category theory such as morphism, functor, natural transformation, and isomorphism are representation of relation. In a category, the object only needs to satisfy the morphism relation in it. Category theorists do not need to point out or care about what constitutes an object and a single object has no significance in a category. The relation between objects is the focus of category theory.

Category theory deals with mathematical concepts in an abstract way, turning these concepts into sets of morphisms. If we have two theories each represented as a category whose object is the theoretical model, the morphism preserves the structure of the model in a suitable sense. If there is an equivalence relation for these categories, we can say that the two theories are categorically equivalent.

Many scientific theories can actually be described as a collection of certain mathematical structures. There are two main reasons for applying the equivalence of category theory in the study of scientific theories. First, the inherent properties of scientific theories requirement, for example, the theory has property of coordinate transformation invariance (isomorphism). Second, the equivalence idea of category theory can simplify theoretical models, so that the model can be greatly simplified and the research work can be achieved easily. The structured thinking way has great revelation and help to master and create new scientific theories.